\documentclass[twocolumn,showpacs,preprintnumbers,amsmath,amssymb,prc]{revtex4}

\usepackage{graphicx}
\usepackage{dcolumn}
\usepackage{bm}
\usepackage{color}

\begin{document}


\title{Understanding the symmetry energy using data from the ALADIN-2000 Collaboration  taken at the GSI Large Neutron Detector }

\author{Sanjeev Kumar}

\author{Y. G. Ma\footnote{Author to whom all correspondence should be addressed:
ygma@sinap.ac.cn}}
\affiliation{Shanghai Institute of Applied
Physics, Chinese Academy of Sciences, Shanghai 201800, China}

\date{\today}

\begin{abstract}
The present study deals with the extraction of 
the
symmetry energy from
heavy-ion collisions
at intermediate energies. Using the isospin quantum
molecular dynamical (IQMD) model, the
dependence of 
the sum of the charge number for fragments 
with
$Z\ge2$
($Z_\textrm{bound}$)
on the
multiplicity of neutrons ($M_n$)
from the projectile spectator fragmentation of $^{124}$Sn
and $^{124}$La at 600 MeV/nucleon is compared with the
experimental results of the ALADIN-2000 Collaboration. The comparison suggest 
a soft symmetry energy. In addition,
the sensitivities of the symmetry energy towards the 
$Z_\textrm{bound}$ dependence on 
proton multiplicity ($M_p$),
neutron-to-proton single [$R(n/p)$] and
double ratio [$R_D(n/p)$] are also examined.
The  $Z_\textrm{bound}$ dependence of
$R(n/p)$ is found to be the most
sensitive observable towards the symmetry energy. The ALADIN Collaboration
should extend the results for $R(n/p)$ in the near future.
\end{abstract}

\maketitle


There is an ongoing effort, within the nuclear physics and nuclear astrophysics communities,
to constrain the density dependence of
the symmetry energy (DDSE) \cite{Latt04,Li08,Qin}. 
Such constraints are also 
urgently needed because of the unique advantage 
of the DDSE in understanding the
nuclear equation of state (NEOS) of asymmetric 
nuclear matter. In the
last couple of years \cite{Li08,Qin,Zhan08, Sun10,Kuma11,Wolt09,Li09,Feng10,Ruso11,Cozm11,Feng12,Trau11},
heavy-ion collisions (HICs) at intermediate energies,
due to their
ability
to cover the 
low-density
as well as high-density region, 
have been
considered an
important tool for the determination of the sub- and supra-saturation density dependence of the
symmetry energy.
At present,
the
high-density behavior of the symmetry energy is 
of considerably greater uncertainty
compared to the low-density 
behavior. 
However, 
a detailed analysis is still
needed at low density 
in order 
to determine the specific characteristics such as the nucleon-nucleon cross section, the symmetry energy coefficient, method of clusterization etc.
 \cite{Zhan08,Sun10,Kuma11,Li97}.
Further progress in this direction 
will depend on the
results from
experiments that are
still ongoing 
at laboratories worldwide.

At present,
the results for
the high-density behavior of the symmetry energy are
interesting but vary drastically 
between models. 
In the following, we will give
a brief overview
of
the theoretical as well as 
the combined theoretical and experimental progress.
Theoretically,
the single and double neutron-to-proton 
ratios
\cite{Li02,Li07},
single and double $\pi^-/\pi^+$
ratios
\cite{Wolt09,Li09, Feng10,Li02,Q05},
$\Sigma^-/\Sigma^+$ ratio \cite{Q05}, $K^-/K^+$ ratio \cite{Wolt09}, and isospin
fractionation \cite{Li07} are 
available
as 
sensitive probes 
to study 
the high-density 
behavior of the symmetry energy. 
In addition, flow parameters are relevant.
Flow in heavy-ion collisions
is studied in 
terms
of the relative and differential
directed flow of neutron-to-proton, $^3$H to
$^3$He  \cite{Li02}, 
ratio and the 
difference \cite{Ruso11,Cozm11} of neutron-to-proton elliptic flow.
Recently, the transverse emission of isospin ratios 
has also been
suggested as a  potential candidate for the determination of the symmetry energy
\cite{Feng12}.

In spite of considerable 
efforts,
it is necessary to
compare
the theory with
experimental findings.
For example,
in 2009,
results from the FOPI collaboration 
for the $\pi^-/\pi^+$ ratio
were
compared with
isospin-dependent
Boltzmann-Uehling-Uhlenbeck (IBUU04) \cite{Li09} and
Lanzhou quantum molecular dynamics (LQMD) \cite{Feng10} model calculations.
The results were in contradiction between the models.
In 2011, 
neutron and proton elliptic flow at 400 MeV/nucleon was
compared with 
findings from 
the FOPI Collaboration 
by two different groups \cite{Ruso11,Cozm11}. One 
group 
predicted the softness of 
the
symmetry
energy with $\gamma_i=0.9$ using the ratio of neutron-to-proton elliptic flow \cite{Ruso11}. 
The other  group
\cite{Cozm11}  also found
evidence 
for a
soft symmetry energy with $x= 2$ using the difference of neutron-to-proton elliptic flow.
However, the approaches for the
determination
of
the symmetry energy were different: in the first study, the symmetry energy was momentum
independent, while in 
the
latter one, it was 
obtained
from momentum-dependent
interactions.
At the end
of 2011 \cite{Trau11}, the multiplicity of neutrons ($M_n$)  
 data from
projectile spectator fragmentation at 600 MeV/nucleon 
at
GSI was compared
with 
statistical multifragmentation model (SMM) calculations. The study
pointed out the essential reduction in the strength of
the
symmetry energy.
Even after the reduction of the symmetry energy coefficient, the data was
reproduced
satisfactorily
only in the range $Z_\textrm{bound}/Z_\textrm{Projectile} \ge 0.5$.

It is clear that 
the
statistical model 
failed to reproduce the data
at low $Z_\textrm{bound}$.
This raises the important question whether a dynamical model
provides an adequate description
because
the mechanism behind the
statistical and dynamical models is
different.
In 
the present
paper, we will try to answer
this question by comparing the above experimental findings with
isospin
quantum molecular dynamical (IQMD) model \cite{Hart98} calculations. In the literature, no one has compared this
data with dynamical model calculations
in order
to extract the symmetry energy.
Success in this regard,
to a certain extent,
may help the community
reduce the uncertainty
in the determination of 
the
symmetry energy.
In addition,
we will also check the
sensitivity of 
the
$Z_\textrm{bound}$ dependence 
on the proton
multiplicity($M_p$), 
and the
neutron-to-proton
single [$R(n/p)$] and double ratios [$R_D(n/p)$] 
with regard to
the symmetry energy.
This will help in providing the most sensitive observable 
for
the symmetry energy.

The study is performed within the framework of IQMD model.
It has been applied successfully for the determination of 
isospin asymmetric nuclear matter equation of state. The isospin effects in the model have been included in terms of the symmetry potential, 
Coulomb potential, and isospin dependent nucleon-nucleon (NN) cross sections.
In the model \cite{Hart98}, nucleons are represented by the wave packets
\cite{Aich91}. The time evolution of
the nucleons, which are under the self-consistency generated mean field, 
is described by the well known Hamilton's equations of motion. The Hamilton $(H)$
is the sum of the kinetic energy and effective interaction potential.
The interaction potential is composed of Coulomb ($V_\textrm{Coul}$),
Yukawa ($V_\textrm{Yukawa}$), local ($V_\textrm{loc}$) and momentum dependent interactions ($V_\textrm{MDI}$).
The  expressions for $V_\textrm{Coul}$ and $V_\textrm{Yukawa}$ have been derived by us and others
in the Refs.~\cite{Hart98}.  The local interaction potential $V_\textrm{loc}$ is
originated from the Skyrme energy density function. On  the basis of this,
the local potential energy density is expanded as:
\begin{equation}
U_\textrm{loc}~=~\frac{\alpha}{2}\frac{\rho^2}{\rho_0}~+~\frac{\beta}{\gamma+1}
\frac{\rho^{\gamma+1}}{\rho_0^{\gamma}}+
E^\textrm{pot}_\textrm{Sym}(\rho)\rho\delta^2,
\label{equation2}
\end{equation}
where $\alpha$, $\beta$, and $\gamma$ are the parametrized values
to specify the particular NEOS. The detailed table of the values is presented in the
Ref.~\cite{Hart98,Aich91}.

The $E^\textrm{pot}_\textrm{Sym}$  is the symmetry potential energy,
which is parametrized using the 
microscopic or phenomenological many body theory calculations. The total symmetry energy per nucleon employed in the simulation
is the sum of the kinetic and potential terms and is given as
\begin{eqnarray}
E_\textrm{Sym}(\rho)&=&\frac{C_\textrm{s,k}}{2}\left(\frac{\rho}{\rho_0}\right)^{2/3}+\frac{C_\textrm{s,p}}{2}\left(\frac{\rho}{\rho_0}\right)^{\gamma_i},
\label{equation3}
\end{eqnarray}
 where $C_\textrm{s,k}~=~ \frac{\hbar^2}{3m}\left(\frac{3\pi^2\rho_0}{2}\right)^{2/3}~\approx~25$
MeV is the symmetry kinetic energy coefficient and $C_\textrm{s,p}~=~35.19~$MeV is the symmetry potential
energy coefficient.

The momentum dependent potential has been implemented from
Ref.\cite{Aich91} and is expressed as following: $V_\textrm{MDI}=C_\textrm{mom}\ln^2[\epsilon(\Delta p)^{2}+1]\frac{\rho}{\rho_0}\delta({r}^\prime -{r})$.
Here $C_\textrm{mom}$ = 1.57 MeV and $\epsilon=5\times10^{-4}\frac{c^2}{MeV^2}$. The momentum
is given in units of MeV/c.

Finally, combining all the potentials, we got an isospin, density and
momentum dependent single particle potential in nuclear matter,
which is written as follow:
\begin{eqnarray}
V_{\tau}(\rho,\delta,p)&&= \alpha\left(\frac{\rho}{\rho_0}\right)+
\beta\left(\frac{\rho}{\rho_0}\right)^{\gamma}+
E^\textrm{pot}_\textrm{Sym}(\rho)\delta^2 \nonumber\\
& &+\frac{\partial E^\textrm{pot}_\textrm{Sym}(\rho)}{\partial\rho}\rho\delta^2+
E^\textrm{pot}_\textrm{Sym}(\rho)\rho\frac{\partial\delta^2}
{\partial\rho_{\tau,\tau^\prime}}\nonumber\\
&&+C_\textrm{mom}\ln^2[\epsilon(\Delta p)^{2}+1]\frac{\rho}{\rho_0}.
\end{eqnarray}
Here $\tau \neq \tau^\prime$, $\frac{\partial\delta^2}{\partial\rho_n}~=~
\frac{4 \delta \rho_p}{\rho^2}$, and $\frac{\partial\delta^2}
{\partial\rho_p}~=~\frac{-4 \delta \rho_n}{\rho^2}$.

In the present simulations, the parameters $\alpha$, $\beta$, and $\gamma$
are -390 MeV, 320 MeV and 1.14, respectively. They give rise  to the soft momentum dependent 
equation of state (SMD). The $\gamma_i$ parameter in Eq.~\ref{equation3} describes the strength of the 
symmetry energy. The $\gamma_i~=~0.5$ and $1.5$
refe to the soft and hard symmetry energies, resepectively. 
Moreover, the isospin and energy dependent NN cross section in the collision
term and the quantum feature in terms of Pauli blocking  
is implemented.  

Following the experimental technique \cite{Trau11}, the neutron and 
proton multiplicity is calculated using the cluster recognization method. In this method,  the particles with relative momentum smaller than $P_\textrm{Fermi}$
and relative distance smaller than $R_{0}$ are coalesced into one
cluster. In the present work, the value of $R_{0}$ and $P_\textrm{Fermi}$ are 3.5 fm and 268 MeV/c, respectively.

In order to perform the analysis,
we have simulated thousands of event for the reactions of $^{124}$Sn+$^{nat}$Sn
and $^{124}$La+$^{nat}$Sn at incident energy 600 MeV/nucleon along the whole
$Z_{bound}$ range. The effect of different targets on the projectile
fragmentation is also observed by simulating  the reaction of $^{124}$Sn+$^{124}$Sn.
In the figures, the horizontal (vertical) bar around the displayed symbols
is the actual $Z_\textrm{bound}$ bin size (error bar as 95$\%$ confidence level).
The 95$\%$ confidence level is $\pm$ Standard Deviation (S.D.). 
The different colored lines with the same colored symbols 
are drawn spline to guide the trend of the different data points as well as reader's eyes.

\begin{figure}
\vspace{-0.3cm}
\includegraphics[width=85mm]{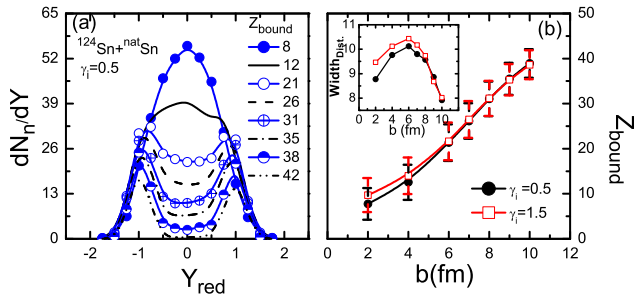}
\vspace{-2.0cm}
\caption{\label{fig:1}(Color online) (a) Rapidity distribution of
neutrons for $^{124}$Sn+$^{nat}$Sn at 600 MeV/nucleon. The different lines and
symbols correspond the values at different $Z_\textrm{bound}$. (b) Impact parameter
dependence of $Z_\textrm{bound}$ with soft (solid circle with black line) and hard (open square with red line) symmetry energy.
The inset is the impact parameter dependence of the width of $Z_\textrm{bound}$ distribution.}
\end{figure}

\begin{figure}[htb]
\vspace{-2.4cm}
\includegraphics[width=85mm]{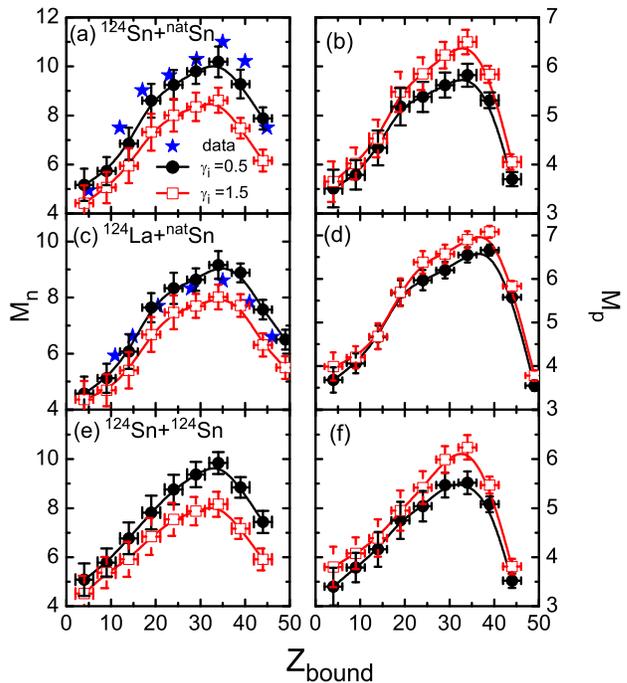}
\vspace{-0.7cm}
\caption{\label{fig:2}(Color online) $Z_\textrm{bound}$ dependence of
$M_n$ ($M_p$) in left (right) panels
from projectile spectator fragmentation for different projectile-target combinations. The open square (solid circle) including lines have the
same meaning as in the (b) part of the Fig. \ref{fig:1}. The experimental data of ALADIN Collaborations is displayed
by blue stars.}
\end{figure}

\begin{figure}
\vspace{-1.6cm}
\includegraphics[width=80mm]{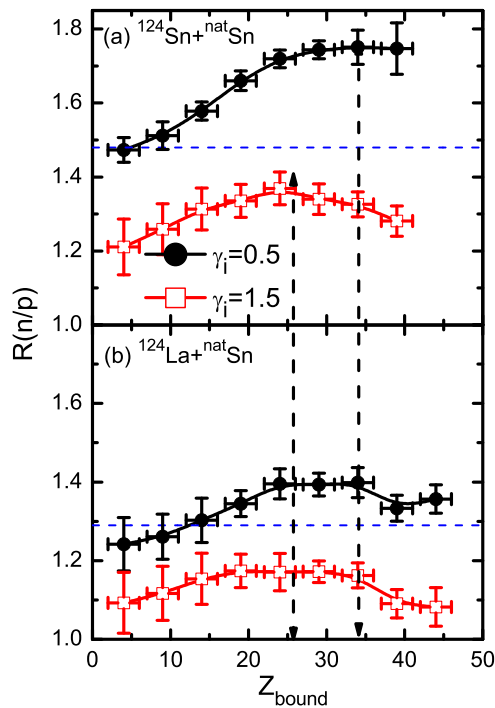}
\vspace{-0.5cm}
\caption{\label{fig:3}(Color online) $Z_\textrm{bound}$ dependence of $R(n/p)$ for neutron-rich (neutron-poor) $^{124}$Sn ($^{124}$La) projectile spectator
fragmentation in top (bottom) panels. The symbol and lines have the same meaning as
in Fig. \ref{fig:2}. The vertical dashed lines correspond to  rough peak positions of $Z_\textrm{bound}$ for $R(n/p)$ with hard and soft symmetry energy, respectively (from left to right).  }
\end{figure}

\begin{figure}
\vspace{-1.4cm}
\includegraphics[width=80mm]{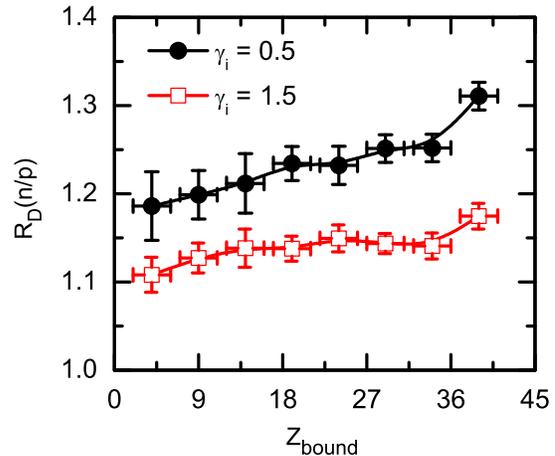}
\vspace{-3.2cm}
\caption{\label{fig:4}(Color online) $Z_\textrm{bound}$ dependence of double neutrons to protons ratio
($R_D(n/p)=\frac{R(n/p)~for~^{124}Sn}{R(n/p)~for~^{124}La}$) with
soft and hard symmetry energy.}
\vspace{0.2 cm}
\end{figure}

\begin{figure}
\vspace{-0.4cm}
\includegraphics[width=90mm]{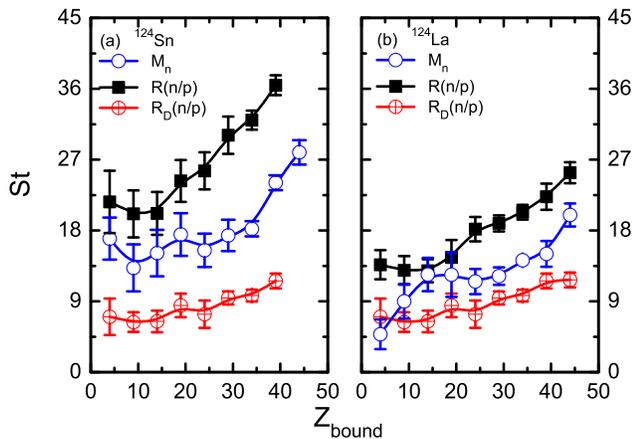}
\caption{\label{fig:5}(Color online) $Z_{bound}$ dependence of sensitivity of
different observables (shown in Fig. \ref{fig:2}, \ref{fig:3}, and \ref{fig:4})
towards the symmetry energy. (a) for  $^{124}$Sn+$^{nat}$Sn and (b) for $^{124}$La+$^{nat}$Sn.
}
\end{figure}

It is important to realize the following experimental filters in the theoretical simulations: (1) the rapidity distribution cut for the projectile spectator
fragmentation region, and (2) the correlation between the $Z_\textrm{bound}$  and impact
parameter (b) \cite{Schu96}. 
For the first factor, in Fig. \ref{fig:1}(a), the
 $Z_\textrm{bound}$ dependence of the rapidity  distribution of neutrons for $^{124}$Sn+$^{nat}$Sn at 600
MeV/nucleon is displayed. At low
$Z_\textrm{bound}$, a single Gaussian shape distribution is observed, which is found to split into two Gaussians
at higher $Z_\textrm{bound}$. From here and with the help of literature \cite{Kuma10}, 
the mid-rapidity region is specified around the zero value of rapidity. It is 
changing into projectile (target)-like matter towards the negative (positive) rapidity values. Using the experimental technique \cite{Trau11}, 
the projectile spectator region in the theoretical calculations  is found to be around -0.9 rapidity value.
 For the second factor,  in  Fig. \ref{fig:1}(b),
the impact parameter dependence of $Z_\textrm{bound}$  is displayed. 
The $Z_\textrm{bound}$ is
calculated at different absolute impact parameters.  
The $Z_{bound}$
increases linearly with the impact parameter. 
It illustrates that the correlation between the two parameters in the dynamical
model agrees with the experimental viewpoint. After the above confirmation, in the rest of the analysis, 
the $Z_\textrm{bound}$ is divided into different bins with the size $\pm2.5$ (Fig. \ref{fig:2}-\ref{fig:5}) by considering the 
triangular distribution of the impact parameter.   
 
In Fig. \ref{fig:1}(b), 
the sensitivity of the symmetry energy on the impact parameter dependence of $Z_\textrm{bound}$ is also checked.
No sensitivity is observed at peripheral geometries, while the visible sensitivity has been found below
semi-central geometries. This provides an indication of  the larger $Z_\textrm{bound}$ with the hard symmetry energy.
It is possible because of different mechanism for $Z_\textrm{bound}$ at central and peripheral geometries \cite{Kuma12}.
To assure that the visible sensitivity of the symmetry energy is  whether or not  due to the statistical error,
the concept of width of $Z_\textrm{bound}$ distribution is used.
The width of distribution is full width at
half maximal (FWHM), which is calculated as $2.35 \times$ S.D..
 The inset in the figure
displays the impact parameter dependence of width of $Z_\textrm{bound}$ distribution with the 
soft and hard symmetry energy.  The symmetry energy sensitivity of width of $Z_\textrm{bound}$ distribution
resembles with the $Z_{bound}$ distribution, indicating sensitivity of impact parameter selector ($Z_{bound}$)
to the symmetry energy is due to the statistical fluctuations. 
This is understandable as $Z_\textrm{bound}$ is a global variable, where the symmetry energy sensitivity becomes
weak. In the following, we need to search for some potential sensitive observables.

In the left panels of Fig. \ref{fig:2}, the $Z_\textrm{bound}$ dependence of  $M_n$
from the projectile spectator decay is displayed.  The first two panels in the figure are
compared with the experimental data of the ALADIN-2000 Collaboration \cite{Trau11}.
Identical to the experimental data, the neutron-rich projectile exhibits a 
maximum of neutrons $M_{n} \approx 11$ for $^{124}$Sn and $M_{n} \approx 9$
for $^{124}$La at the value of $Z_\textrm{bound}/Z_\textrm{Projectile} \approx 0.7$ in the presence of the
soft symmetry energy. Above this $Z_\textrm{bound}$, more neutrons are 
 bound inside fragments and hence  the multiplicity of neutrons  decreases.

Interestingly, the multiplicity of neutrons presents sensitivity to the
strength of the symmetry energy. In all the panels, the soft symmetry energy yields
more neutrons compared to the hard symmetry energy. The sensitivity towards the symmetry energy has been seen more
clearly from the more neutron-rich projectile $^{124}$Sn compared to the neutron-poor
$^{124}$La. The previous comparison with the SMM calculations reproduced the data satisfactorily
only in the range $Z_\textrm{bound}/Z_\textrm{Projectile} \ge 0.5$. The present IQMD model reproduces the experimental data along
the whole $Z_\textrm{bound}$ range. However, some discrepancies are observed for neutron-poor projectile
$^{124}$La compared to neutron-rich projectile $^{124}$Sn. This depends on the different strength of the symmetry energy
and Coulomb interactions in different projectiles. Owing to the more neutron-rich nature of $^{124}$Sn, stronger symmetry energy effect has
been observed and hence a good agreement with the data is obtained. While in case of $^{124}$La, due to the presence of 4 extra protons
compared to $^{124}$Sn, the
contribution from Coulomb interactions results in the deviation of the calculations from the
experimental data.  The enhanced production of protons in Fig. 2(d) compared to
Fig. 2(b) is further supporting the discrepancies.  

In addition, the multiplicity of neutrons from the
projectile spectator fragmentation is strongly dependent on the
type of projectile [(a) and (c)], while, is weakly dependent on the type of the
target [(a) and (e)]. The related study has also been conducted at relatively low energy 25 MeV/nucleon for
$^{40}$Ca$~+~^{40,48}$Ca, $^{46}$Ti reactions at the 4$\pi$ CHIMERA detector \cite{Amor09}. They showed that competition between
fusion-like and binary reactions in selected centrality bin can constrain the parametrization of the symmetry energy.

In the right panels of Fig. \ref{fig:2}, the $Z_\textrm{bound}$ dependence of  proton multiplicity $M_p$
 from the projectile spectator decay is displayed.
 The behavior of the curves resembles with the left panels.
 The differences
between the left and right panels are as follows: (1) The $M_p$ is more
with the hard symmetry energy, which was true with the soft for the $M_n$.
This is because of the repulsive (attractive) nature of symmetry energy for neutrons (protons).
At the freeze-out time, the soft symmetry energy has high magnitude, which will lead to  
the production of neutrons. It has alternatively led to the less production of protons;
(2) More protons are produced for neutron-poor projectile $^{124}$La compared to the
neutron-rich projectile $^{124}$Sn; (3) The sensitivity of symmetry energy
is weak on the  $M_p$ in $^{124}$La projectile, which is due to the dominance of Coulomb interactions. These factors are revealing the importance of the
symmetry energy for neutron-rich reactions.

Now we will analyze the $Z_\textrm{bound}$ dependence of neutron-to-proton single ratio $R(n/p)$ from the projectile spectator decay.  In Fig. \ref{fig:3},
 the $R(n/p)$ is
displayed for the $^{124}$Sn+$^{nat}$Sn (top panel) and
$^{124}$La+$^{nat}$Sn (bottom panel) reactions. The single ratio is found to be more sensitive
towards the soft symmetry energy. Particularly, the $R(n/p)$ with the soft (hard) symmetry
energy is higher (lower) than the ratio of system [shown by blue horizontal line]. 
This is attributable to the large (small) magnitude of soft (hard) symmetry energy at freeze-out time. 
In addition, Figs. \ref{fig:2} and \ref{fig:3} has some similarities and differences. 
The behavior of the single ratio curves  
almost resemble with the $Z_\textrm{bound}$ dependence of $M_n$ and $M_p$, i.e. it also exhibits a maximum  or saturates with $Z_\textrm{bound}$.  The difference is that the maximal value for $R(n/p)$
is noticed at different $Z_\textrm{bound}/Z_\textrm{projectile}$ with the soft (at 0.7) and hard (at 0.5) symmetry energy, which was at the
same  $Z_\textrm{bound}/Z_{projectile}$ for the maximal $M_n$ and $M_p$ (at $\approx 0.7$). 
It means that the single ratio has more (less) neutron-rich effects along $Z_\textrm{bound}$ with
the soft (hard) symmetry energy. This is once again due to  different strength of symmetry energy and Coulomb interactions originated from the soft and 
stiff parametrization of the symmetry energy.    

As discussed in the Ref.~\cite{Zhan08}, the double neutron-to-proton
ratio $R_D(n/p)$ from the two isotopes of a system is the most sensitive
observable for the determination of the symmetry energy. 
In the present study, the projectiles are isobars and not the isotopes.
The
$Z_\textrm{bound}$ dependence of $R_D(n/p)$ using the isobars is shown in Fig. \ref{fig:4}.
From the comparison of Fig. \ref{fig:3} with Fig. \ref{fig:4},  it is found that the $Z_\textrm{bound}$ dependence of the single as well as double ratio is  
more sensitive to the soft symmetry
energy.  In contrary, the double ratio, comparing to the   
single ratio, has relatively weak sensitivity  toward the symmetry energy along the whole $Z_\textrm{bound}$ region.  
Moreover, the $R_D(n/p)$ with the  hard symmetry energy is weakly dependent on the $Z_\textrm{bound}$. This may be the effect of four
extra protons in the $^{124}$La projectile, where the contribution from the 
Coulomb interactions can not be canceled, even by taking the double ratio.

Finally, to declare the most sensitive observable toward the symmetry energy among the
all three, i.e. $M_n$, $R(n/p)$, and $R_D(n/p)$, we 
have defined the sensitivity factor ($St$) as follow 
\begin{equation}
St =\frac{|X_{\gamma_i=0.5}-X_{\gamma_i=1.5}|}{X_{\gamma_i=1.5}} \times 100
\end{equation}
where $X$ = $M_n$, $R(n/p)$ or $R_D(n/p)$.

From the results of $St$ (Fig. \ref{fig:5}), we can see that the $St$ values for all the parameters are more sensitive for neutron-rich projectile, illustrating that neutron-rich nucleus is  a good candidate for the
extraction of the symmetry energy.
As was expected from Fig. \ref{fig:4}, the $St$ factor of $R_D(n/p)$ shows  the least sensitivity to the symmetry energy 
in this particular situation of different isobars. Between the other two,  i.e. $M_n$
and $R(n/p)$, the $R(n/p)$ shows more sensitivity  to
the symmetry energy.
This sensitivity becomes more stronger
with the increase of $Z_\textrm{bound}$.  Therefore, it is advisable that experimental Collaboration should also make the data for the $R(n/p)$ available, which 
can help to check the validity of our results for $R(n/p)$ and $R_D(n/p)$. 

In conclusion, using the IQMD model, the ALADIN-2000 Collaboration data for multiplicity of neutrons extracted from the  projectile spectator
fragmentation is
satisfactorily reproduced with the soft symmetry energy. Along with multiplicity
of neutrons, proton multiplicity has also a peak around
$Z_\textrm{bound}/Z_\textrm{porjectile}\approx0.7$ with the soft as well as the hard
symmetry energy. Interestingly, the $R(n/p)$ produced a peak
around  $Z_\textrm{bound}/Z_\textrm{porjectile}\approx0.7$ (0.5) with the soft
(hard) symmetry energy. The $R(n/p)$ for the different isobar
projectiles is the most sensitive observable towards
the symmetry energy in comparison to multiplicity as well as $R_D(n/p)$.
In the near future, we hope experimental Collaborations  make the results for proton multiplicity and $R(n/p)$ available.

This work is  supported in part by the Major State Basic Research Development Program of China under Contract No. 2013CB834405,
 the Chinese Academy of Sciences
 Program for young international scientists under the Grant
No. 2010Y2JB02, the  NSFC  of China under contract No.
11035009 and No, 10979074,  and the Knowledge Innovation Project of the
Chinese Academy of Sciences under Grant No. KJCX2-EW-N01.

\end{document}